\documentclass[12pt]{iopart}
\pdfoutput=1

\expandafter\let\csname equation*\endcsname\relax
\expandafter\let\csname endequation*\endcsname\relax
\usepackage{amsmath}

\usepackage[utf8]{inputenc}
\usepackage[T1]{fontenc}
\usepackage{amssymb, amsfonts}
\usepackage{mathrsfs}
\usepackage{bm}
\usepackage{graphicx}
\usepackage{xcolor}
\usepackage{float}

\newcommand{\be}{\begin{equation}}
\newcommand{\ee}{\end{equation}}

\begin{document}

\title[Competition and coexistence of superconducting symmetries in $p$-wave magnets]{Competition and coexistence of superconducting symmetries in $p$-wave magnets}

\author{J E C Carmelo$^{1,2}$, I R Pimentel$^{1}$ and P D Sacramento$^{2}$}

\address{$^1$ Departamento de F\'{\i}sica and CFTC, Faculdade de Ci\^encias, Universidade de Lisboa, Campo Grande, 1749-016 Lisboa, Portugal}
\address{$^2$ CeFEMA, Instituto Superior T\'{e}cnico, Universidade de Lisboa, Av. Rovisco Pais, 1049-001 Lisboa, Portugal}
\date{\today}

\begin{abstract}

We investigate the interplay between unconventional magnetism and superconductivity in a model of a $p$-wave magnet on a square lattice. Using a self-consistent Bogoliubov-de-Gennes approach, we analyze the pairing amplitudes, competition, and coexistence of spin-singlet $s$-wave and spin-triplet $p$-wave pairings in the presence of a magnetic texture with a helical structure along the $x$ direction that is repeated in the $y$ direction.

We find that the magnetic helix selectively stabilizes different pairing symmetries depending on its orientation and strength. In particular, mixed-spin $p_x$-wave pairing is enhanced at intermediate magnetic couplings and equal-spin $p_y$-wave pairing is robust and insensitive to all coupling intensities. When the multiple order parameters are simultaneously considered, we find regimes of coexistence and competition. Increasing the magnetic coupling drives two quantum phase transitions. The first from dominant spin-singlet $s$-wave to mixed-spin triplet $p_x$-wave pairings in a regime of coexistence. The second from spin-singlet $s$-wave and mixed-spin triplet $p$-wave pairings with total spin projection $S_z=0$ to dominant equal-spin triplet $p_y$-wave pairings with $S_z=\pm1$ in a regime of mutually exclusive superconducting phases.

Our results demonstrate that $p$-wave magnetic order does not merely diminish spin-singlet $s$-wave superconductivity but can actively promote and stabilize spin-triplet $p$-wave pairing, both intrinsically and in proximity to spin-singlet $s$-wave superconductors. These findings highlight unconventional magnets as promising materials for realizing robust triplet superconductivity.

\end{abstract}


\maketitle

\section{Introduction}

The discovery of unconventional magnets (UMs), specifically $d$-wave altermagnets and $p$-wave magnets, has recently redefined the classification of magnetic phases \cite{Smejkal, Mazin, Hellenes2024, Brekke2024, Mineev2026}. These materials are characterized by a vanishing net magnetization despite the presence of non-relativistic spin-split electronic bands. Altermagnets exhibit collinear magnetic moments and even-parity spin-split bands while breaking time-reversal symmetry and preserving a combined symmetry $[C_{2}||C_{4z}\boldsymbol{t}]$ \cite{Smejkal}. In turn, $p$-wave magnets exhibit non-collinear, coplanar magnetization and odd-parity spin-split bands protected by a composite time-reversal and translation symmetry $[C_{2\perp}|\mathcal{T}\boldsymbol{t}]$ \cite{Hellenes2024}. $p$-wave magnets have very high TMR and support non-relativistic Edelstein and spin-galvanic effects. Experimental platforms for such textures include materials like $\text{CeNiAsO}$, $\text{Mn}_3\text{GaN}$.
Helical structures have in general attracted considerable interest. Examples are chiral metals or
chiral molecules, for
instance in the context of chiral induced selective spin transport. Molecules
such as DNA or similar structures, either single stranded or doubly
stranded, have been studied
\cite{xie,guo1,gutierrez1,eremko,gutierrez2,gutierrez3,naaman,varela,caetano,matityahu,utsumi,inui,evers}.
A question that may arise is a possible topological nature of the systems considered \cite{Madeira}. 

The interplay between unconventional magnetism and superconductivity is a burgeoning frontier 
for realizing exotic phases \cite{Khodas2026, Maeda2025, Fukaya2025, Li2025, ChenLu2025, Sukhachov2025, Bobkov2025, PHFu2025, PHFu2026}. 
While magnetism is typically viewed as antagonistic to superconductivity, 
several systems show complex phase diagrams where competition or even coexistence between magnetic phases
and superconducting phases is found, such as in heavy-fermion systems 
\cite{Monthoux,Pfleiderer,Grosche,Knebel,Sacramento}, cuprates \cite{Damascelli, Lee}, 
iron-based materials \cite{Zhao,Luetkens} and metallic ferromagnets \cite{Brando}.
Recent research suggests that unconventional magnetism can provide a route to robust pairing, being strong candidates to host or induce both singlet and triplet superconductivity due to their symmetries and lack of significant stray fields. While altermagnets suppress singlet $s$-wave superconductivity, short-range altermagnetic fluctuations have been shown to enhance singlet $d$-wave pairing \cite{Li2025}. Conversely, $p$-wave magnets like helimagnets can host singlet $s$-wave superconductivity with a critical temperature that is gradually diminished as magnetic couplings are increased \cite{Sukhachov2025}.  Moreover, both altermagnets and $p$-wave magnets exhibit momentum-orbital locking that favors inter-orbital spin-triplet states \cite{Fukaya2025, ChenLu2025}. For instance, in $p$-wave magnets the possibility of triplet $p$-wave superconductivity is reported, in particular in mixed-spin channels of even frequency \cite{Maeda2025, Sukhachov2025}.  This is because the orbital symmetry of the $p$-wave magnet can match that of the $p$-wave pairing, with spin-splitting that can favor triplet pairing. They have also been proposed to host Cooper pairs that exist as a 50:50 mixture of singlet and triplet states of enhanced resilience against pair-breaking \cite{Khodas2026}. This makes $p$-wave magnets very interesting tools to potentially induce robust $p$-wave superconductivity, including exotic and topological phases 
\cite{Ezawa2024_pwave_SC}.

The proximity effect in magnetic-superconducting heterostructures provides a fertile setting for realizing these states at the interface layer. Recent studies by Bobkov et al. \cite{Bobkov2025} have demonstrated that atomically thin $s$-wave superconductors-helimagnet ($p$-wave magnetic) heterostructures host proximity-induced $p$-wave triplet correlations that can carry non-dissipative transport spin currents. Similarly, Fukaya et al. \cite{Fukaya2026}, Maeda et al. \cite{Maeda2025} and Salehi et al. \cite{Salehi2025} have shown that $p$-wave magnets can convert singlet into mixed spin-triplet $p$-wave correlations.

To date, the scientific community has primarily explored these systems through two lenses. First, extensive phenomenological classifications that have established how UMs act as spin-mixers, converting singlets to triplets while transferring their parity to the superconducting order parameter \cite{Maeda2025, Khodas2026}. Second, a vast body of literature focuses on proximity effects in heterostructures, where superconductivity is externally imposed \cite{Bobkov2025, Fukaya2026, Salehi2025, Alam2025, BoFu2025, BoLu2026, Zhang2025, Kayatz2026}. These studies have identified signatures such as zero-energy flat bands, transverse spin supercurrents, and Bogoliubov Fermi surfaces \cite{Fukaya2026, Salehi2025, BoFu2025, BoLu2026}. 

For these reasons, it is interesting to study singlet $s$-wave and triplet $p$-wave superconductivity in or in proximity to $p$-wave magnets in some detail. Much of the existing literature relies on proximity models where the superconducting gap is assumed to be a constant or a self-energy, rather than being solved self-consistently within the magnetic background. Additionally, while induced triplet amplitudes are well-documented \cite{Maeda2025, Bobkov2025}, the self-consistent stability of these pairings, as well as their competition with singlet superconductivity remain under-explored in this picture.

Our work focuses on providing a self-consistent microscopic solution for superconductivity on a square lattice choosing a four-site unit cell forming a helical spin structure along the $x$ direction that is repeated along the $y$ direction to model a $p$-wave magnet ($p_x$-wave magnet). By solving the Bogoliubov-de-Gennes equations in the presence of a magnetic texture that is modeled by a local helical Zeeman-like field, we demonstrate that $p$-wave magnets do not merely host induced pairing but can actively stabilize and enhance triplet $p$-wave components at moderate to high magnetic couplings, something which does not occur for $s$-wave superconductivity. We predict that this can also occur in the interface of $s$-wave superconductors and $p$-wave magnets, in agreement with \cite{Maeda2025, Bobkov2025, Fukaya2026}. We show that the symmetry of $p$-wave magnetic helices selectively amplifies distinct triplet pairing channels depending on the plane orientation of the magnetic helix and the magnetic coupling intensity. When the $s$-wave and $p$-wave order parameters are introduced simultaneously, this leads to a phase transition from spin-singlet/mixed-spin triplet pairing to equal-spin triplet superconductivity beyond a critical magnetic coupling. Furthermore, the coexistence between $s$-wave and mixed-spin $p$-wave superconductivity in certain regimes can increase their robustness to the magnetic coupling, exhibiting higher critical coupling.

These results provide potential ways to not only induce but also control robust superconducting phases with strong $s$-wave or $p$-wave superconductivity, tunable through the magnetic coupling, magnetic orientations and heterostructure engineering of $p$-wave magnets and superconductors. 

The paper is organized as follows:
In Section 2, a model for a $p$-wave magnet with superconductivity is set-up and the formalism is outlined. In Section 3, the results are presented. 
In Subsection 3.1, we present zero temperature phase diagrams of $s$-wave and $p$-wave pairings in different spin channels as a function of the magnetic and superconducting couplings.
In Subsection 3.2, the temperature is varied in conjunction with the magnetic coupling to estimate critical magnetic couplings for each symmetry.
In Subsection 3.3, we display how pairing amplitudes of $s$-wave and different $p$-wave symmetries depend on the magnetic coupling when they compete, and also the case where an $s$-wave gap is imposed externally is presented, simulating the interface between a $s$-wave superconductor and a $p$-wave magnet.
In Section 4 the conclusions are presented.

\section{Model for a $p$-wave magnet with superconductivity}

\begin{figure}
\centering
(a)\includegraphics[width=6.5cm]{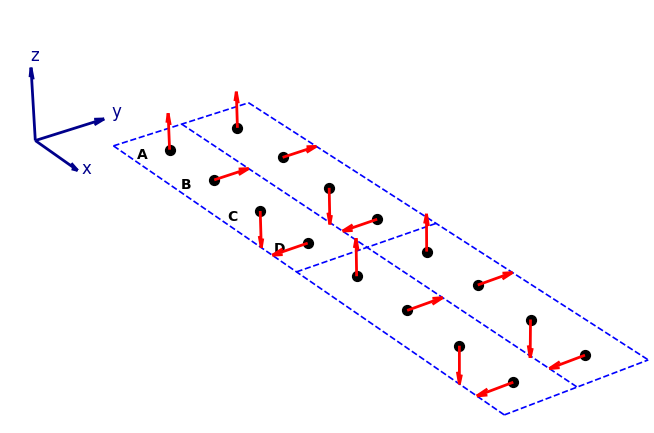} 
(b)\includegraphics[width=6.5cm]{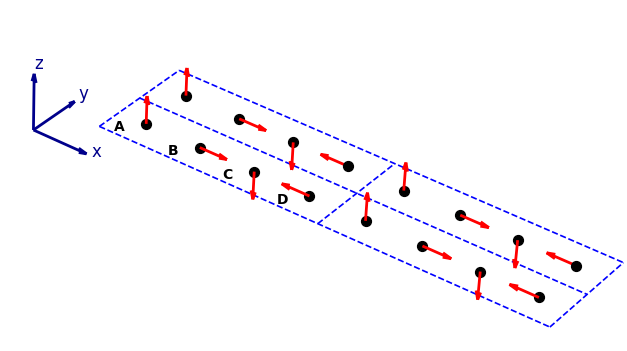} 
\caption{$p$-wave magnet on a square lattice. The helical structure is along the $x$ direction. Each unit cell has 4 sites: A, B, C, D. 
(a) Helix with spins in the $zy$ plane ($pM_{zy}$).
(b) Helix with spins in the $zx$ plane ($pM_{zx}$).}
\label{fig2}
\end{figure}

We consider a system with $p$-wave magnetism and superconductivity, on a square lattice, described by the Hamiltonian

\begin{align}
\mathcal{H} &=-t \sum_{\langle i,j \rangle, \sigma} c^\dagger_{i\sigma} c_{j\sigma}
 - J \sum_{i, \sigma, \sigma'} \mathbf{M}_i \cdot c^\dagger_{i\sigma} \boldsymbol{\sigma}_{\sigma\sigma'} c_{i\sigma'} 
   - \mu \sum_{i,\sigma} c^\dagger_{i\sigma} c_{i\sigma} \nonumber\\
  &\quad + \sum_{\langle i,j \rangle,\sigma,\sigma'}  (\Delta_{i,j,\sigma,\sigma'} c^\dagger_{i\sigma} c^\dagger_{j\sigma'} + \text{h.c.} )\label{eq:1}
\end{align}
where  $c^\dagger_{i\sigma}, c_{j\sigma'}$ are creation and annihilation electron operators on sites $i$ and $j$ with spin $\sigma$ and $\sigma'$, respectively. $t$ is the hopping parameter and $\mu$ is the chemical potential. To model the $p$-wave magnet, we consider the spin structures shown in Fig. \ref{fig2} where we have unit cells with 4 sites along the $x$ direction of the lattice, labeled $\text{A,B,C,D}$, each with a local Zeeman-like field $J\mathbf{M}_{i}$ acting on the electron on site $i$ as $J\mathbf{M}_i \cdot  \boldsymbol{\sigma}$, where $\boldsymbol{\sigma}$ represents the Pauli vector, $\mathbf{M}_i$ is a unit vector and $J$ is the magnetic coupling intensity. The configuration of the magnetic helix is specified by $J \mathbf{M}_{i}$, the local magnetization vectors at the sublattices. An equivalent way to model a helimagnet is to consider a homogeneous ferromagnet with spin orbit coupling in the form of complex hoppings in the $x$ direction \cite{Brekke2024}. In this work, a spin period of rotation of 4$a$ is considered as an example, where $a$ is the lattice constant, other choices for the helix pitch lead to similar results. 

We consider two choices for the plane of orientation of the spins in the helix. In the  $pM_{zy}$ configuration the spins are in the $zy$-plane, transverse to the helix,
\begin{equation}
    pM_{zy}:
    \quad \mathbf{M}_A = (0,0,1), \;
    \mathbf{M}_B = (0,1,0), \;
    \mathbf{M}_C = (0,0,-1), \;
    \mathbf{M}_D = (0,-1,0).
\end{equation}
On the other hand, in the $pM_{zx}$ configuration the spins are in the $zx$-plane, transverse or parallel to the helix,
\begin{equation}
   pM_{zx}:
    \quad \mathbf{M}_A = (0,0,1), \;
    \mathbf{M}_B = (1,0,0), \;
    \mathbf{M}_C = (0,0,-1), \;
    \mathbf{M}_D = (-1,0,0).
\end{equation}
In the $y$ direction the system is made up of copies of the helix aligned in the $x$ direction. 

Such spin helices are known to generate a set of energy bands with spin density distributions that are odd-parity in the momentum along the $x$ direction, with the preservation of a composite time-reversal and translation symmetry, characteristic of a $p$-wave magnet. 

We consider superconductivity in a mean-field approximation. The superconducting gap functions can be written in real space, connecting two sites and spin components as
\begin{equation}
\Delta_{i,j,\sigma,\sigma'}=
V_{i,j,\sigma,\sigma'}
\left\langle 
c_{i,\sigma} \, c_{j,\sigma'} 
\right\rangle.
\end{equation}
The superconducting couplings $V_{i,j,\sigma,\sigma'}>0$ describe an effective attractive interaction between electrons, favoring their pairing. The pairings satisfy fermionic antisymmetry
$\Delta_{i,j,\sigma,\sigma'} = -\Delta_{j,i,\sigma',\sigma}$, which constrains the allowed combinations of spatial and spin symmetries. The spatial structure of the interactions determines the symmetry of the superconducting order parameter. In particular, onsite interactions denoted by $V_s$ favor spin-singlet $s$-wave pairing, while nearest-neighbor interactions denoted by $V_p$ give rise to spin-triplet $p$-wave pairing components. 

We will look for the zero temperature values and the temperature dependence
of the order parameters of the various spatial symmetries shown in Table \ref{table:oprs}.

\begin{table}[h!]
\caption{Real-space expressions of order parameters for the basic superconductivity symmetries in terms of the spatial relation of the gaps for sites given by $i$ and $\hat{\mathbf{x}}$ and $\hat{\mathbf{y}}$ are the vectors connecting nearest-neighbor sites along $x$ and $y$, respectively.}

\centering
\begin{tabular}{|c|c|}
\hline
SC symmetry & Real-space order parameter \\ \hline
spin-singlet $s$-wave& $\Delta_i$ \\ \hline
spin-triplet $p_x$-wave& $\Delta_{i,i+\hat{\mathbf{x}}} - \Delta_{i,i-\hat{\mathbf{x}}}$ \\ \hline
spin-triplet $p_y$-wave& $\Delta_{i,i+\hat{\mathbf{y}}} - \Delta_{i,i-\hat{\mathbf{y}}}$ \\\hline
\end{tabular}
\label{table:oprs}
\end{table}

\begin{figure}
\centering
(a)\includegraphics[width=6.5cm]{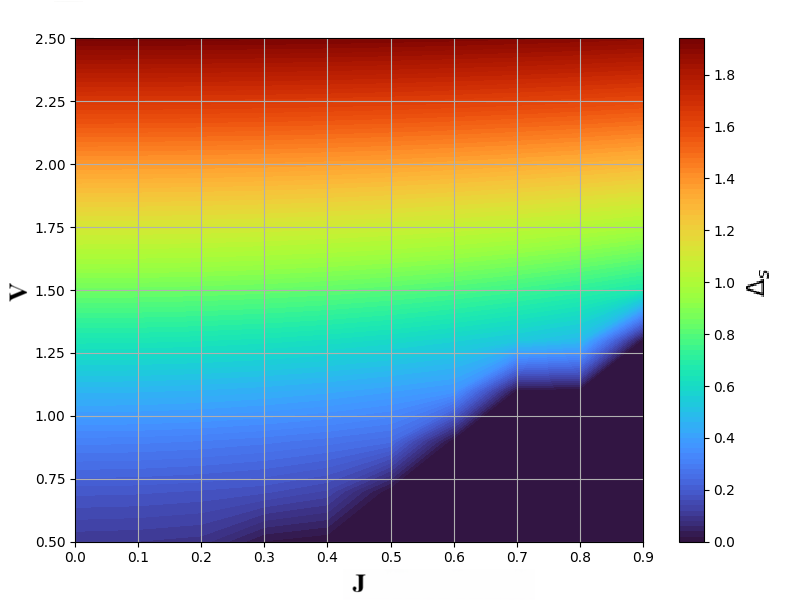} 
(b)\includegraphics[width=6.5cm]{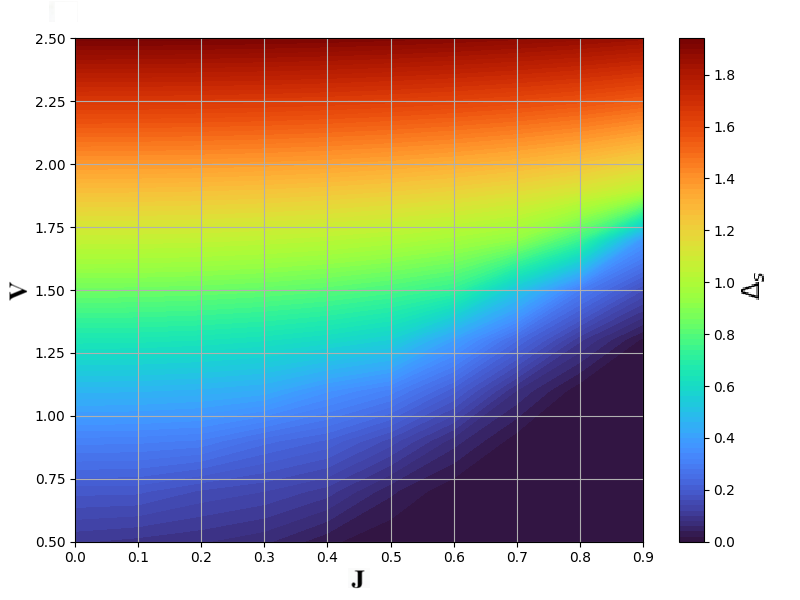} 
\caption{
Spin-singlet $s$-wave order parameter $\Delta_s$ as a function of the magnetic coupling $J$ and the superconducting coupling $V_s$,
(a) for the $pM_{zy}$ plane orientation and (b) for the $pM_{zx}$ plane orientation.}
\label{fig21}
\end{figure}

The order parameters then average the absolute value of the gaps that are needed to define each symmetry over the spatial extent of the system,

\begin{equation}
\Delta_{s} = \frac{1}{N}\sum_{i}
\big|\Delta_{i,i,\uparrow\downarrow}
- \Delta_{i,i,\downarrow\uparrow}\big|,
\label{eq:3s}
\end{equation}

\begin{equation}
\Delta^{\text{mixed-spin}}_{p_{\xi}} = \frac{1}{N}\sum_{i}\sum_{\sigma\neq\sigma'}
\big| \Delta_{i,i+\boldsymbol{\xi},\sigma,\sigma'}
- \Delta_{i,i-\boldsymbol{\xi},\sigma,\sigma'} \big|,
\label{eq:3pm}
\end{equation}

\begin{equation}
\Delta^{\text{equal-spin}}_{p_{\xi}} = \frac{1}{N}\sum_{i, \sigma}
\big| \Delta_{i,i+\boldsymbol{\xi},\sigma,\sigma}
- \Delta_{i,i-\boldsymbol{\xi},\sigma,\sigma} \big|,
\label{eq:3pe}
\end{equation} 
Here $N = 2N_x N_y$, where $N_x$ and $N_y$ denote the number of sites along each direction, and $\boldsymbol{\xi} = \hat{\mathbf{x}}$ or $\hat{\mathbf{y}}$. These definitions provide a measure of $p_x$-wave and $p_y$-wave pairing amplitudes for equal-spin ($\uparrow\uparrow$, $\downarrow\downarrow$) and mixed-spin ($\uparrow\downarrow$, $\downarrow\uparrow$) channels. 

The superconducting order parameters are obtained from the converged solutions to the Bogoliubov-de-Gennes equations under different choices for the magnetic parameters. The calculations are performed with periodic boundary conditions for systems with $N_x=N_y=24$, and choose $\mu=0$ and $t=1$. In the following, we consider both single-channel cases, where only one interaction is non-zero, and coexistence regimes where multiple interaction channels are simultaneously present. 

\section{Results}
\subsection{Phase diagrams for different pairing symmetries}
\begin{figure}
\centering

(a)\includegraphics[width=6cm]{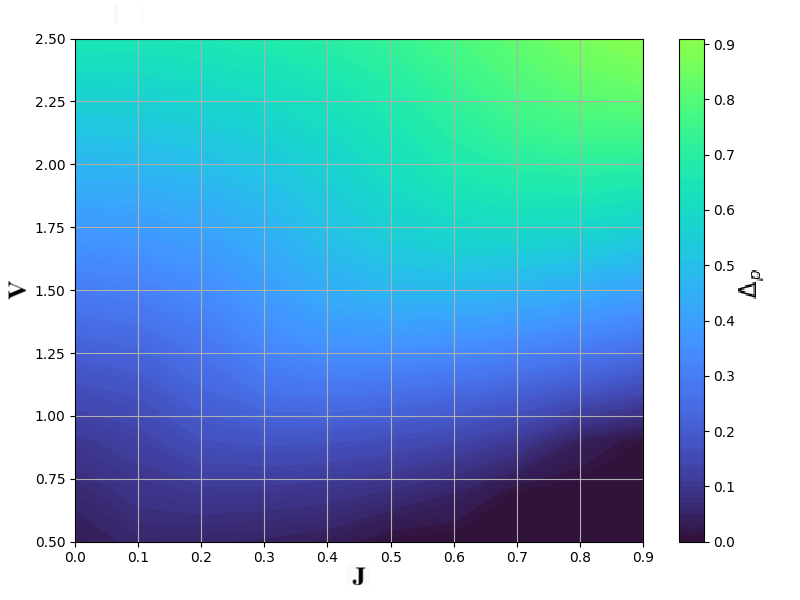} 
(b)\includegraphics[width=6cm]{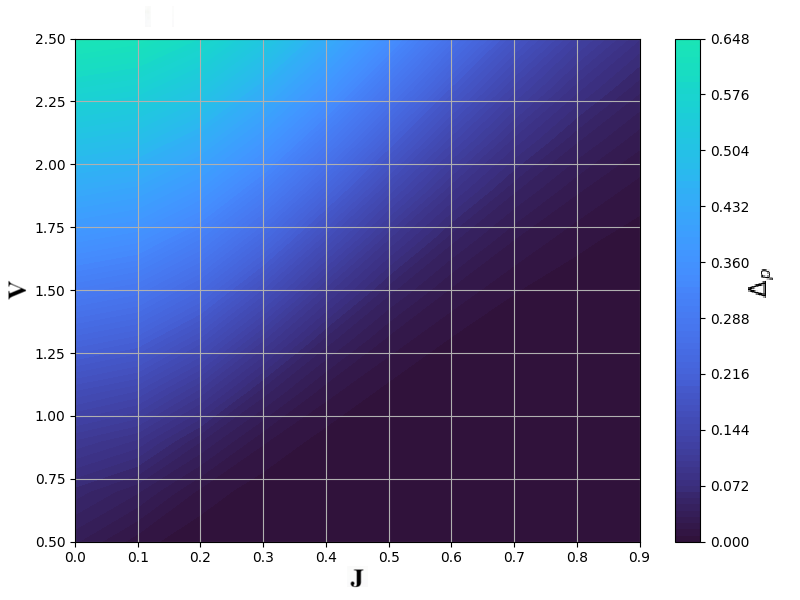} 

(c)\includegraphics[width=6cm]{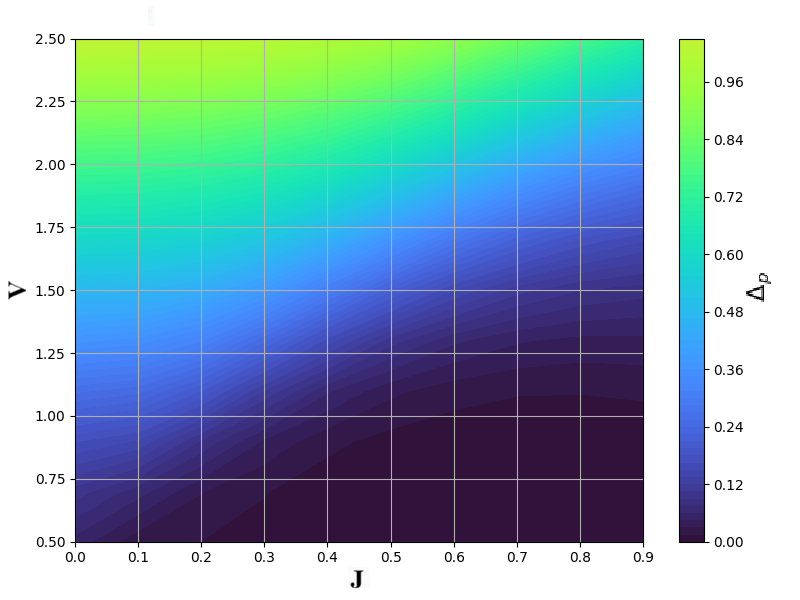} 
(d)\includegraphics[width=6cm]{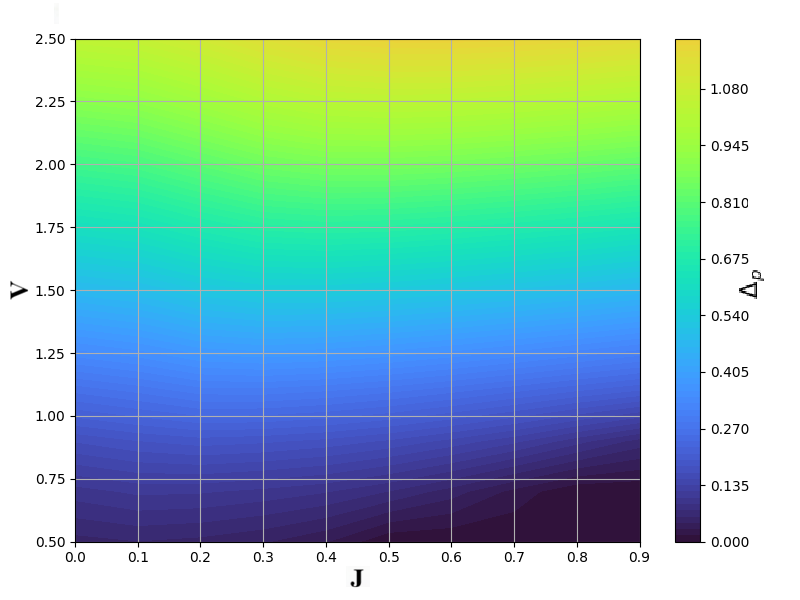} 

(e)\includegraphics[width=6cm]{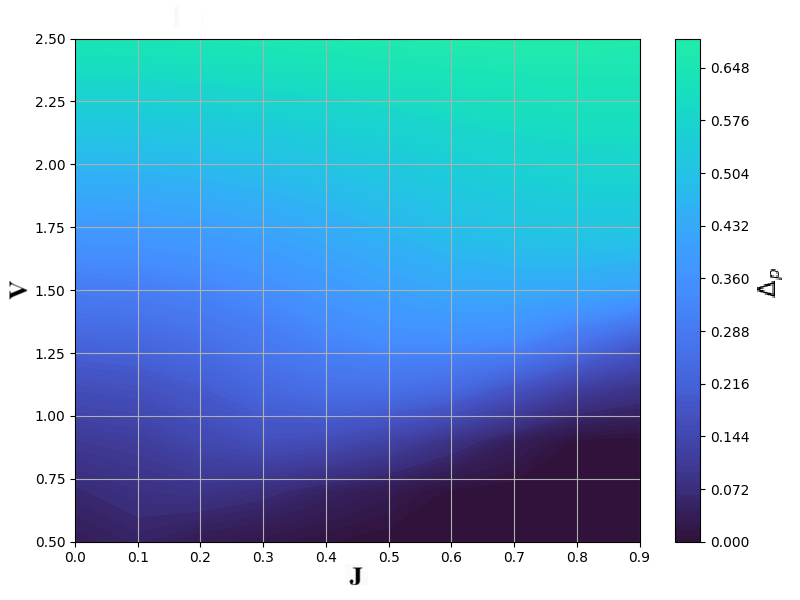} 
(f)\includegraphics[width=6cm]{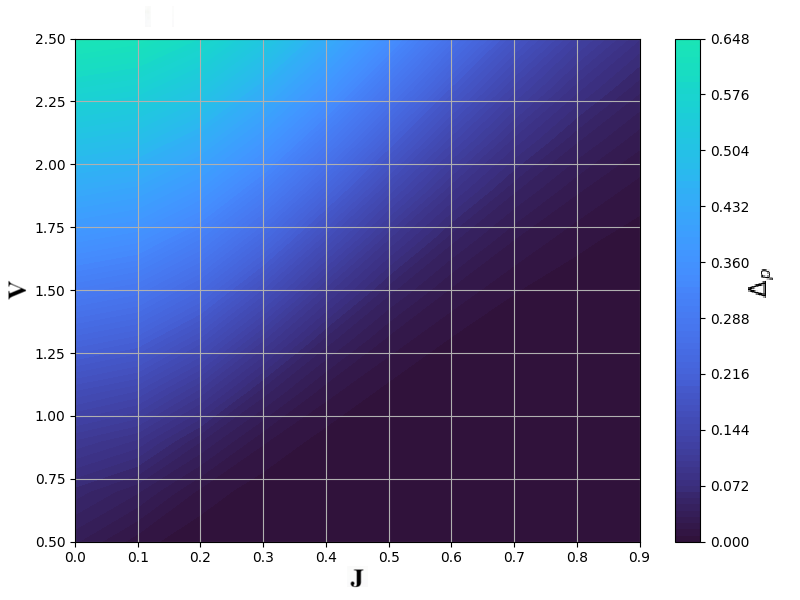} 

(g)\includegraphics[width=6cm]{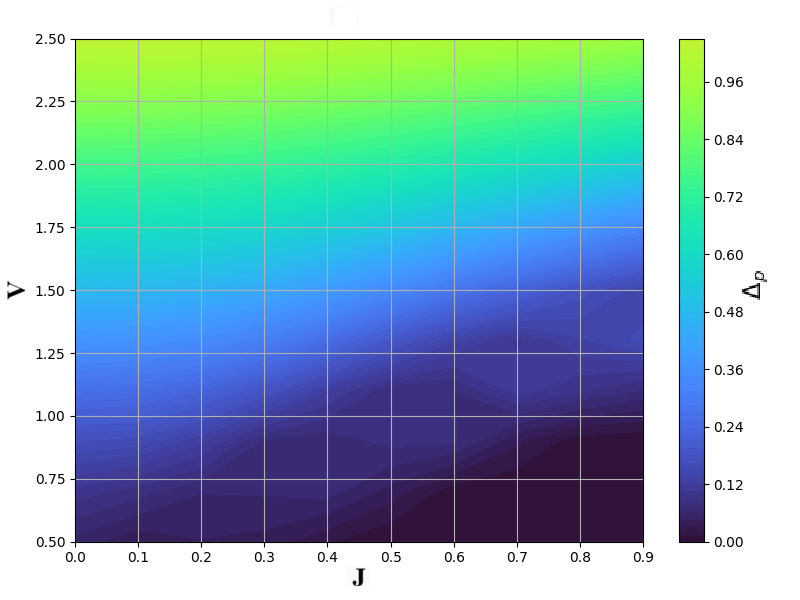} 
(h)\includegraphics[width=6cm]{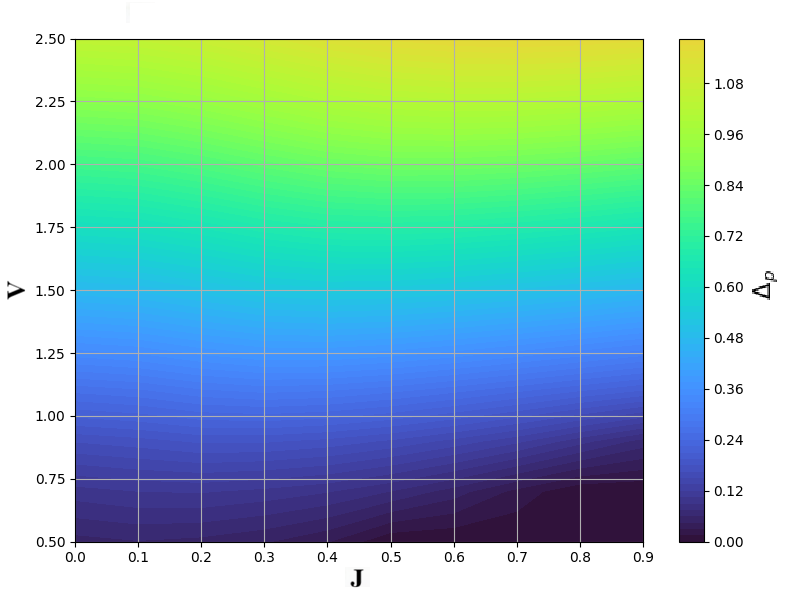} 

\caption{
Spin-triplet $p$-wave order parameter $\Delta_p$ as a function of the magnetic coupling $J$ and the superconducting coupling $V_{p_x},V_{p_y}$ in the $pM_{zy}$ plane orientation for
(a) mixed-spin $p_x$-wave, (b) mixed-spin $p_y$-wave, (c) equal-spin $p_x$-wave, (d) equal-spin $p_y$-wave, and  in the $pM_{zx}$ plane orientation for (e) mixed-spin $p_x$-wave, (f) mixed-spin $p_y$-wave, (g) equal-spin $p_x$-wave, (h) equal-spin $p_y$-wave.}
\label{fig22}
\end{figure}

 The phase diagrams for the order parameters are obtained at zero temperature for each individual symmetry by varying the magnetic coupling $J$ and the corresponding superconducting coupling $V$ (setting all others to zero).

For $s$-wave pairing, we obtained the gap as shown in Fig. \ref{fig21}. The results are in agreement with \cite{Sukhachov2025}, with a pairing amplitude that is large for low to moderate magnetic coupling, with a gap that is continually diminished as $J$ increases, becoming completely suppressed by $J\geq J_c$. While the results are similar for both orientations of the spins in the helix, the $pM_{zy}$ exhibits a sharper, less gradual transition to the non-superconducting state.

For $p$-wave pairings, along the helix $p_x$-wave and perpendicularly to it $p_y$-wave are considered in both equal-spin and mixed-spin channels. The results are displayed in Fig. \ref{fig22} for the different plane orientations.
Note that $p$-wave pairing is non-vanishing and not small even though the $s$-wave pairing seems to be dominant. In addition, as the $s$-wave gaps decline with higher magnetic couplings, some channels of $p$-wave superconductivity remain steady or are even enhanced, surpassing the $s$-wave order parameter in some phase diagram regions. This effect occurs in both planes of orientation, with more striking visibility in the $pM_{zy}$, where the $p$-wave magnetic textures favor mixed-spin $p_x$-wave pairing along the direction of the helix, being the most visibly optimized channel, with a peak at a moderate, non-zero magnetic coupling. Additionally, the equal-spin $p_y$-wave pairings are robust under magnetic intensity, maintaining stable pairing amplitudes as the magnetic coupling is increased. 
On the other hand, equal-spin $p_x$-wave pairings are gradually diminished with increasing $J$, while mixed-spin $p_y$-wave is altogether forbidden in the magnetic regime. 

These findings are in line with the symmetries of the magnetic textures, which constrain and select the symmetries of the superconducting pairings, with spin-splitting patterns and an orbital symmetry that favors the formation of Cooper pairs in a triplet state of the given channels. For instance, while the parallel alignment of spins along the $y$ direction favor an energetic preference for equal-spin pairings, a spin helix with time-reversal symmetry along the $x$ direction favors mixed-spin pairings, something which is in line with predictions found in \cite{Sukhachov2025}. Other orientations of the helix would lead to anisotropic modifications, particularly in the relative strength of $p_x$-wave and $p_y$-wave pairing components, but do not alter the qualitative behavior discussed here.

You can find an outline of the magnetic regimes and their dominant pairing symmetries on Table \ref{table:results_summary_single}.

\begin{table}[h]
\centering
\caption{\label{table:results_summary_single}
Dominant pairing symmetries for $V=1$ considering single order parameter regimes in $J$, obtained from Figs.~\ref{fig21} and \ref{fig22}. }
\begin{tabular}{|c|c|c|}
\hline
Magnetic Orientation 
& $J\lesssim0.6$& $J\gtrsim0.6$\\
\hline
$pM_{zy}, pM_{zx}$& $s$-wave 
& \begin{tabular}[c]{@{}c@{}}
mixed-spin $p_x$-wave \\ 
equal-spin $p_y$-wave
\end{tabular} \\\hline
\end{tabular}
\end{table}

In the following subsections, we will show regimes where the tendency of dominant $s$-wave superconductivity is more clearly reversed, providing further evidence for stable $p$-wave pairing.

\subsection{Temperature effects}

\begin{figure}[htbp]
\centering
(a)\includegraphics[width=5.5cm]{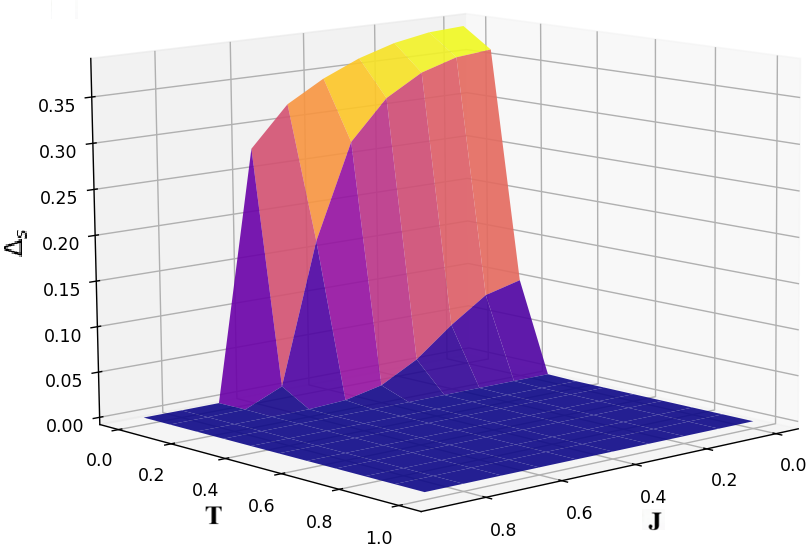} \\[0.5cm] 

(b)\includegraphics[width=5.5cm]{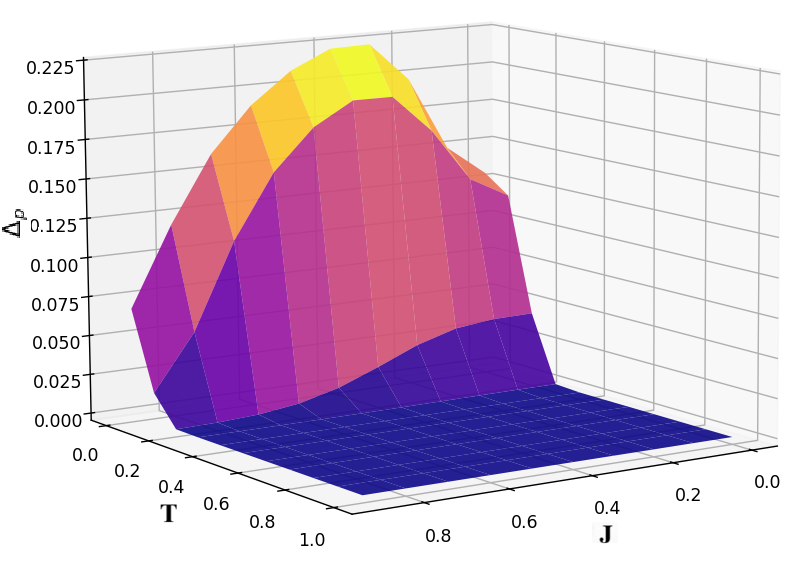} \quad
(c)\includegraphics[width=5.5cm]{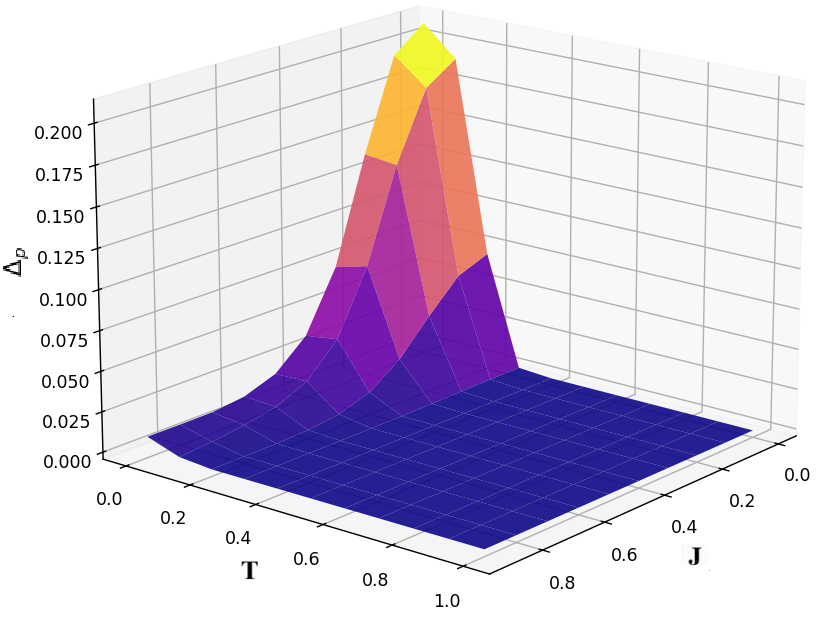} 

\caption{For the $pM_{zy}$ plane orientation, the order parameter as a function of temperature $T$ 
and magnetic coupling $J$ is shown:
(a) $s$-wave superconductivity,
(b) mixed-spin $p_x$-wave superconductivity, and
(c) equal-spin $p_x$-wave superconductivity.}
\label{fig23}
\end{figure}

Besides analyzing the behavior of the individual zero temperature gap functions, we may study their
behavior as a function of temperature, as shown in Fig.\ref{fig23}, obtaining an estimate of the critical temperature $T_c$ of each individual pairing symmetry. Moreover, by fixing the temperature we may estimate the value of the critical
magnetic coupling $J_c$ for which the self-consistent solution of the gap functions vanishes, signaling the transition to the superconducting regime. The zero temperature gap function amplitude 
$\Delta=\Delta(V,J)$ , the critical temperature $T_c=T_c(V,J)$ and the critical magnetic coupling $J_c=J_c(V,T)$ may be compared across superconducting symmetries to determine the dominant pairings of the system. These quantities are analyzed by selecting self-consistent solutions for each type of symmetry separately, with a fixed superconducting coupling of $V=1$ for that corresponding symmetry. We consider the $pM_{zy}$ configuration given that it provides the strongest zero temperature enhancements of $p$-wave superconductivity.

\begin{figure}
\centering

(a)\includegraphics[width=3.5cm]{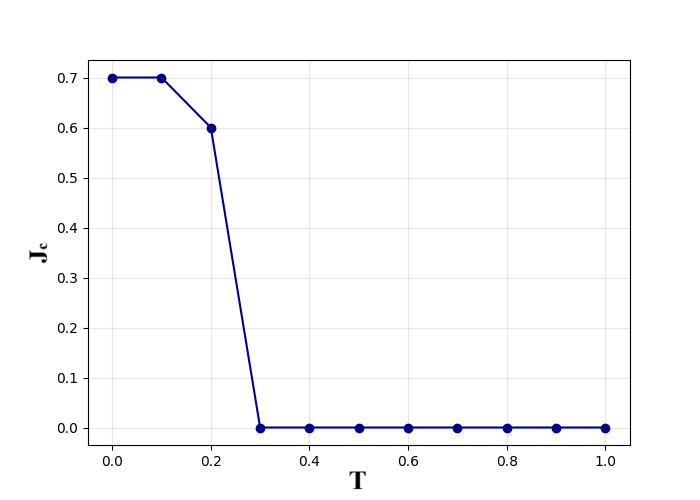} 
(b)\includegraphics[width=3.5cm]{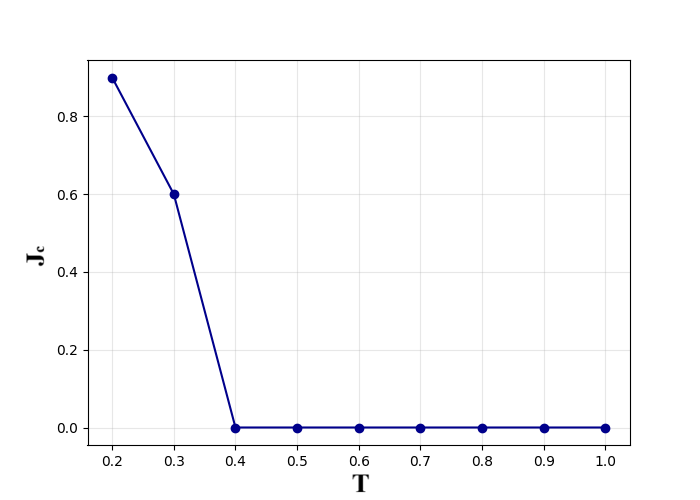} 
(c)\includegraphics[width=3.5cm]{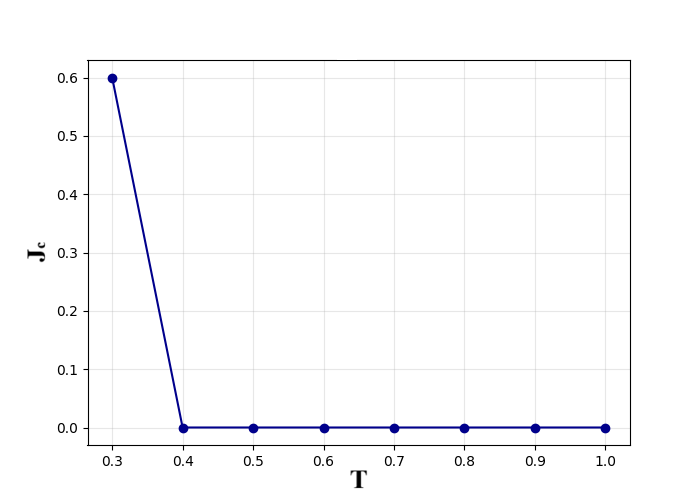} 

\caption{For the $pM_{zy}$ plane orientation, $J_{c}$ of the order parameter as a function of $T$ for
(a) $s$-wave superconductivity.
(b) $p_x$-wave equal-spin, (c) $p_x$-wave mixed-spin.}
\label{fig24}
\end{figure}

In the nonmagnetic limit $J=0$, all pairing symmetries yield $T_c \approx 0.4$. Increasing $J$ suppresses the $T_c$ for all pairing symmetries but with varying degrees of sharpness. For the $s$-wave pairing the $T_c$ vanishes for $J \gtrsim 0.8$, while the equal-spin $p_x$-wave pairing persists with $T_c \approx 0.2$ and the mixed-spin $p_x$-wave pairing with $T_c\approx0.3$ at least up to $J = 0.9$, not vanishing in the considered domain. 

In Fig. \ref{fig24} a focus is placed on the temperature region in which $J_c$ vanishes. Specifically, the $s$-wave gap is fully suppressed with $J_c = 0$ for $T \approx 0.3$, whereas $p_x$-wave pairing channels require a higher temperature of $T \approx 0.4$, to reach the same condition. The mixed-spin $p_x$-wave pairings exhibit greater robustness under thermal and magnetic effects, having $J_c>0.9$ for $T\leq0.2$.

\subsection{Coexistence and competition of different pairing symmetries}

Furthermore, it is interesting to test the simultaneous presence of multiple pairing symmetries with the
$p$-wave magnetic texture at zero temperature. The symmetry of
$p$-wave magnets has been shown to enforce a 50:50 singlet-triplet mixture as the only allowed Cooper pair state \cite{Khodas2026}, motivating the question of how singlet and triplet channels compete and stabilize when treated as independently self-consistent order parameters.  We do this by setting two or more of the coupling strengths, $V_s$, $V_{p_x}$, $V_{p_y}$ equal to 1, allowing them to evolve self-consistently.

Our results reveal a distinction between mixed-spin triplet pairings $S_z=0$ which exhibit cooperative coexistence with spin-singlet $s$-wave pairings, and equal-spin triplet pairings $S_z=\pm1$, which strongly compete with the former pairings, leading to their mutual exclusion in their combined phase diagrams.

\subsubsection{Coexistence and competition of $s$-wave and $p$-wave pairings }

\begin{figure}
\centering

(a)\includegraphics[width=6.5cm]{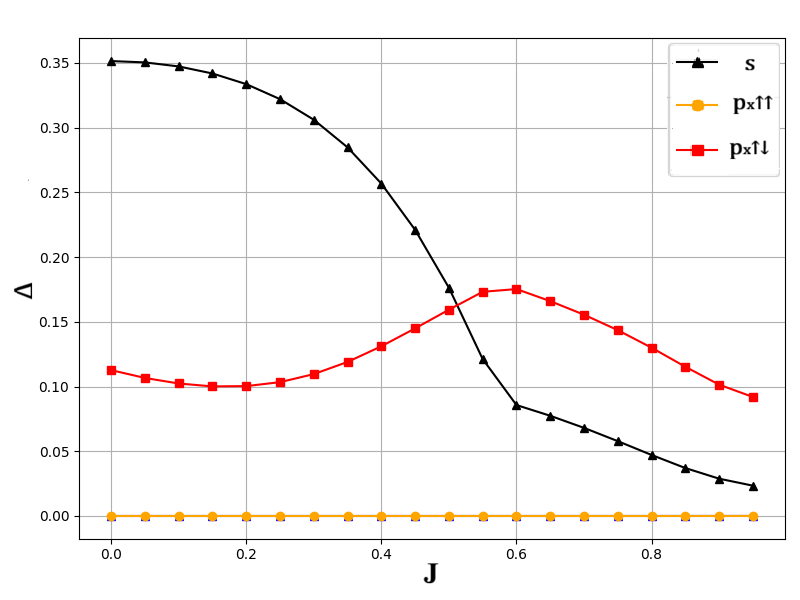} 
(b)\includegraphics[width=6.5cm]{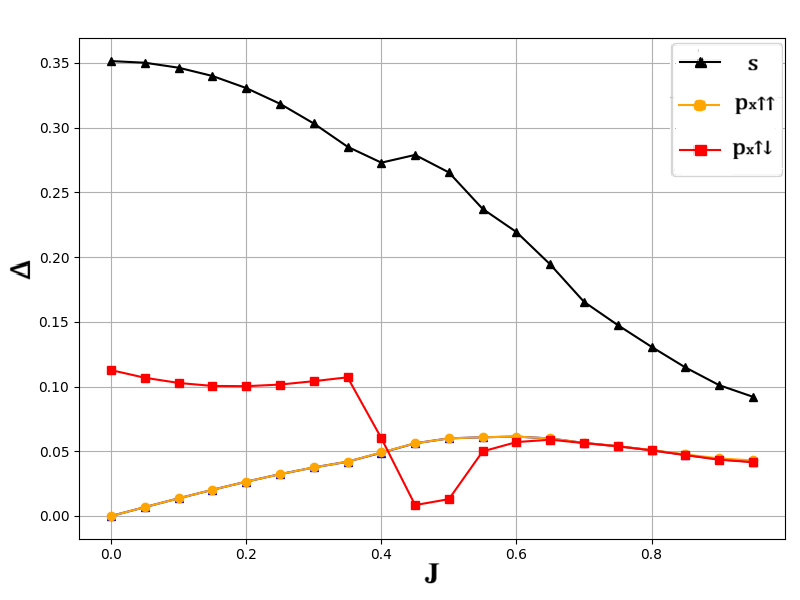} 

(c)\includegraphics[width=6.5cm]{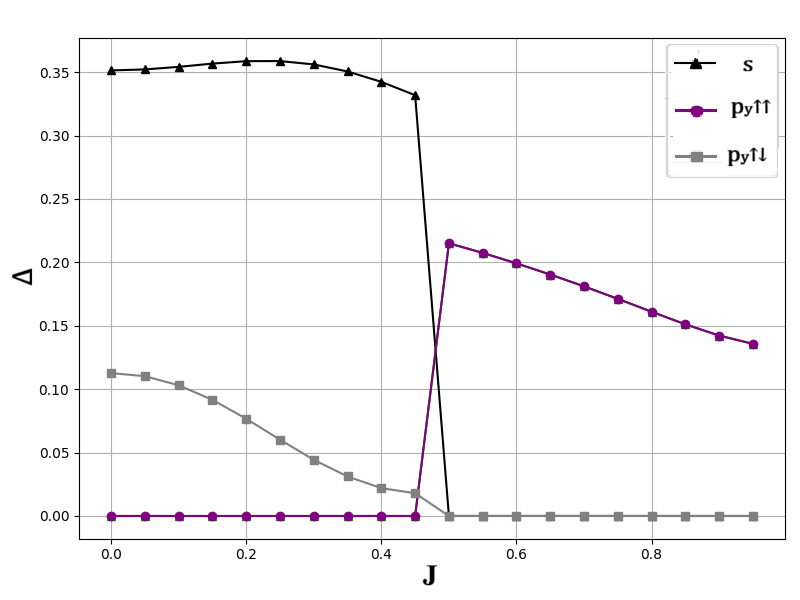} 
(d)\includegraphics[width=6.5cm]{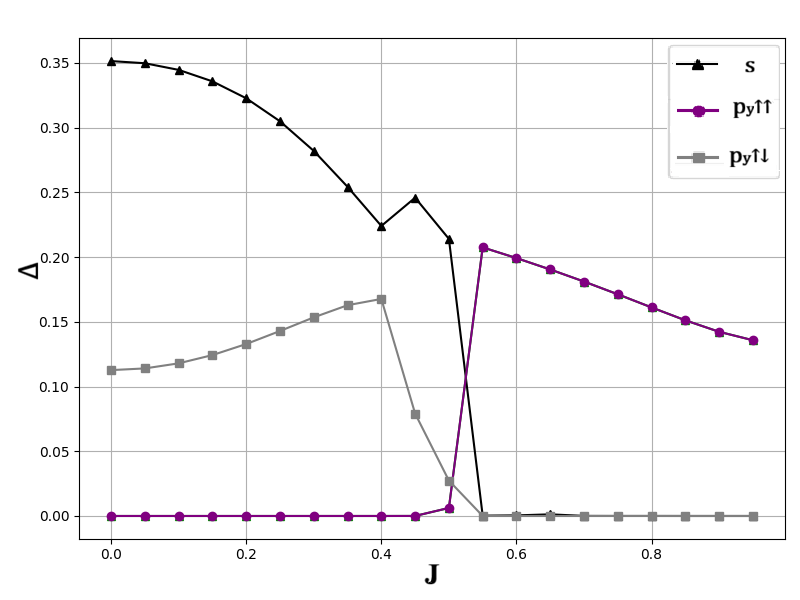} 

\caption{
 Competition between $s$-wave and $p_x$-wave superconductivity (a) in the $pM_{zy}$ plane orientation (b) in the $pM_{zx}$ plane orientation.
Competition between $s$-wave and $p_y$-wave superconductivity (c) in the $pM_{zy}$ plane orientation (d) in the $pM_{zx}$ plane orientation. For the $pM_{zy}$ orientation we find a quantum phase transition for $J=J_{QPT_1}\approx0.5$, that separates a phase with dominant $s$-wave from a phase with dominant mixed-spin $p_x$-wave at higher magnetic couplings. Additionally, for both the $pM_{zy}$ and the $pM_{zx}$ orientations, we find a quantum phase transition for $J=J_{QPT_2}\approx 0.5$, that separates a phase with coexisting $s$-wave and mixed-spin $p_y$-wave from a phase exclusively with equal-spin $p_y$-wave.}
\label{fig25}
\end{figure}

In Fig. \ref{fig25} this is considered for $s$-wave paring and $p_x$-wave pairing or $s$-wave pairing and $p_y$-wave pairing, probing how these spin-triplet channels 'interact' with spin-singlet superconductivity. In the labels we use $s$ for $s$-wave, $p_x\uparrow \uparrow$ or $p_y\uparrow\uparrow$ for equal-spin $p_x$-wave or $p_y$-wave, and $p_x \uparrow\downarrow$ or $p_y \uparrow\downarrow$ for mixed-spin $p_x$-wave or $p_y$-wave pairings, respectively.

For the $pM_{zy}$ orientation, we find that $s$-wave pairing dominates at low magnetic coupling $J$. However, the mixed-spin $p_x$-wave channel is remarkably resilient. As $J$ increases, the singlet s-wave pairing begins to decline, reaching an inflection point where a quantum phase transition takes place $J_{QPT_1} \approx 0.5$ and the mixed-spin $p_x$-wave pairing surpasses it. Despite this, their coexistence leads to a mutual enhancement of robustness for higher magnetic couplings; the critical coupling $J_c$ for both symmetries is higher than when either is solved in isolation. We attribute this to a ``buffer'' effect, where the mixed-spin triplet helps the system accommodate the magnetic splitting that would otherwise be purely pair-breaking for a singlet.

In the $pM_{zx}$ orientation, this coexistence persists, though $s$-wave remains dominant across the sampled range ($J \leq 0.9$), with the convergence of both $p_x$-wave pairing channels' amplitudes to a similar value.

The interaction between $s$-wave and $p_y$-wave symmetries is qualitatively different. For $J < 0.5$, the $s$-wave pairing completely suppresses the equal-spin $p_y$-wave channel, coexisting with the mixed-spin pairing instead. As $J$ approaches a critical threshold, $J_{QPT_2} \approx 0.5$ for both $pM_{zy}$ and $pM_{zx}$ orientations, the system undergoes a sharp quantum phase transition. The $s$-wave and mixed-spin $p_y$-wave amplitudes vanish abruptly, replaced with emerging equal-spin $p_y$-wave superconductivity. This signals a transition from a phase where the Cooper pairs have a total spin projection of $S_z=0$ to one where $S_z=\pm1$ . 

\subsubsection{Full Competition ($s, p_x, p_y$)}  

\begin{figure}
\centering

(a)\includegraphics[width=6.5cm]{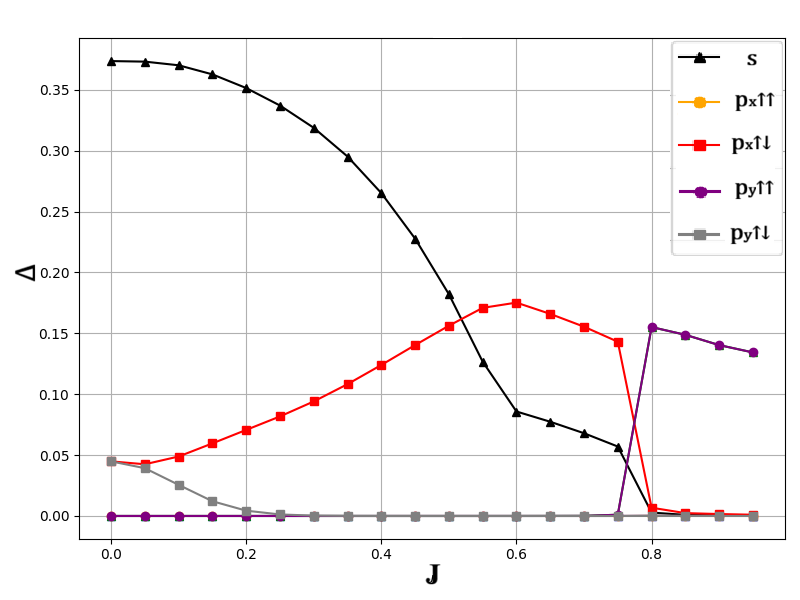} 
(b)\includegraphics[width=6.5cm]{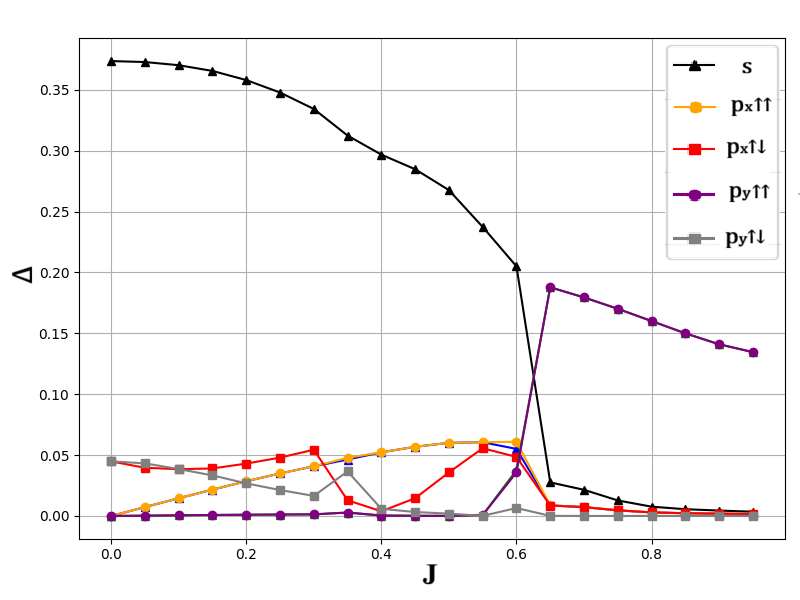} 

\caption{
Competition between $s$-wave, $p_x$-wave and $p_y$-wave superconductivity (a) in the $pM_{zy}$ plane orientation, (b) in the $pM_{zx}$ plane orientation.  For the $pM_{zy}$ orientation equal-spin $p_x$-wave pairings are always zero and we find two quantum phase transitions, the first from dominant $s$-wave to dominant mixed-spin $p_x$-wave at $J= J_{QPT_1}\approx0.5$, and the second from dominant mixed-spin $p_x$-wave to exclusively equal-spin $p_y$-wave pairings at $J= J_{QPT_2}\approx0.8$.  In the $pM_{zx}$ orientations, we find a quantum phase transition for $J=J_{QPT_2}\approx 0.6$, that separates a phase with dominant $s$-wave from a phase exclusively with equal-spin $p_y$-wave pairings.}

\label{fig26a}
\end{figure}

In Fig. \ref{fig26a} we test for the simultaneous presence of all pairings, solving self-consistently for all gap function symmetries.

When all superconducting couplings are simultaneously nonzero ($V_s=V_{p_x}=V_{p_y}=1$), the solutions display a sequence of phases characterized by distinct dominant symmetries as the magnetic coupling is increased. In the low $J$ limit the system is characterized by singlet $s$-wave dominance, with nonzero mixed-spin $p_x$-wave and $p_y$-wave. As the intermediate magnetic regime is approached from below, the mixed-spin $p_x$-wave is increased, overtaking and suppressing the mixed-spin $p_y$-wave that was present at this stage in Fig. \ref{fig25}. Then, in the $pM_{zy}$ configuration the mixed-spin $p_x$-wave becomes the dominant order parameter, surpassing the $s$-wave pairings at $J_{QPT_1}\approx0.5$. Meanwhile in the $pM_{zx}$  orientation, both $p_x$-wave channels converge to similar amplitudes again, failing to surpass $s$-wave superconductivity. Because the introduction of mixed-spin $p_x$-wave results in an increase of the $J_c$ of $s$-wave superconductivity, it delays the quantum phase transition to equal-spin $p_y$-wave pairings, occurring at a higher coupling of $J_{QPT_2}\approx 0.8$ in the $pM_{zy}$ and $J_{QPT_2}\approx 0.6$ in the $pM_{zx}$ orientation. This showcases how $s$-wave and mixed-spin symmetries can strengthen each other as they compete with equal-spin $p_y$-wave symmetries.

\subsubsection{Proximity-Induced Effects and $p$-wave spin-triplet stabilization}

\begin{figure}
\centering

(a)\includegraphics[width=6.5cm]{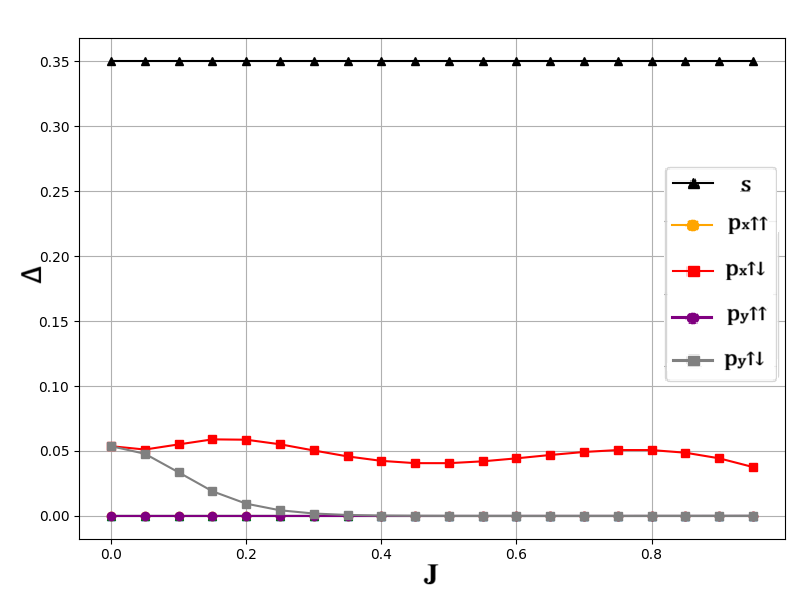} 
(b)\includegraphics[width=6.5cm]{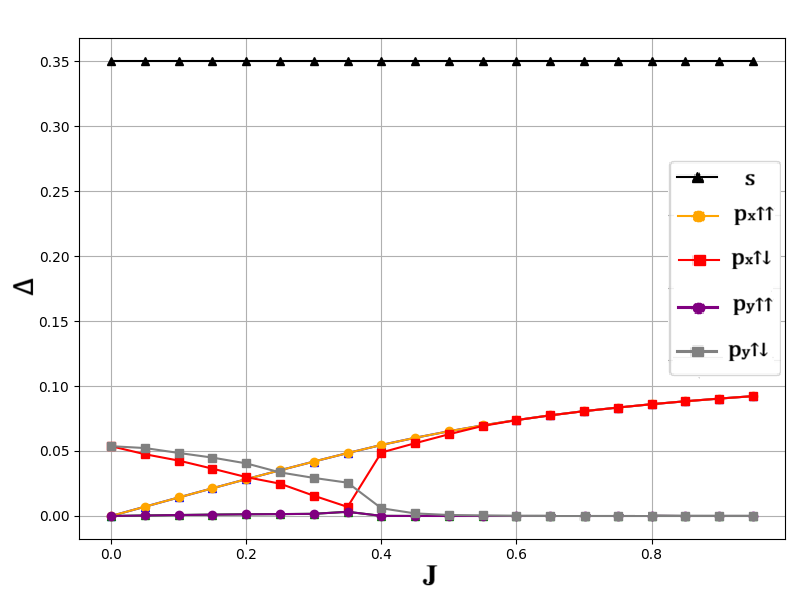} 

\caption{
Proximity-induced competition with fixed $s$-wave gap (a) in the $pM_{zy}$ plane orientation, (b) in the $pM_{zx}$ plane orientation.
}
\label{fig26b}
\end{figure}

To simulate a heterostructure interface, we fix a non-zero $s$-wave gap representing a proximate conventional superconductor and solve for the triplet components self-consistently as shown in Fig. \ref{fig26b}. The presence of a fixed singlet background acts as a symmetry filter. On one hand, the equal-spin $p_y$-wave triplets are completely extinguished for all $J$ values due to the competition described previously, while the same occurs for equal-spin $p_x$-wave for the $pM_{zy}$ plane orientation. On the other hand, the $p_x$-wave triplet channels, including mixed-spin in $pM_{zy}$ and both mixed and equal-spin components in $pM_{zx}$, remain remarkably robust. In the $pM_{zx}$ orientation specifically, the $p_x$-wave order parameters show no signs of decay as $J$ increases within the studied range. This suggests that the proximity-induced $s$-wave gap ``softens'' the pair-breaking effects of the magnetic background, effectively hindering the transition to a pure equal-spin $p_y$-wave state. This synergy allows for the survival of $p_x$-wave triplets at magnetic couplings beyond their $J_c$ in the former setting.

\section{Conclusions}

In this work, we have provided a self-consistent microscopic analysis of the interplay between $p$-wave magnetic textures of helimagnets and $s$-wave and $p$-wave superconductivity on a square lattice. By solving the Bogoliubov-de Gennes equations for various pairing symmetries over a helical magnetic texture, we demonstrated that $p$-wave magnets do not merely act as pair-breakers but can actively stabilize and enhance specific $p$-wave superconducting channels.

A simple analysis of the phase diagrams of individual superconducting order parameters reveals that while $s$-wave superconductivity is dominant for most parameter values, getting diminished and depleted by higher magnetic couplings alone, equal-spin $p_y$-wave is insensitive to such magnetic coupling increases, and mixed-spin $p_x$-wave displays enhancements followed by diminishment in some regions. An analysis of the thermal effects on the order parameters reveals mixed-spin $p_x$-wave is more robust than $s$-wave superconductivity, displaying higher $T_c$ at nonzero $J$ and larger $J_c$ for a wide temperature window.

A further consideration of how the order parameters compete when introduced simultaneously reveals the existence of both coexistence and competition between symmetries. The main findings are displayed on Tables 3,4 and 5.

\begin{table}[h]
\centering
\caption{\label{table:results_summary_competition}
Dominant pairing symmetries considering the coexistence and competition regimes between singlet $s$-wave and triplet $p_x$-wave superconductivity for $V=1$ in $J$, obtained from 
Fig.~\ref{fig25}}

\begin{tabular}{|c|c|c|}
\hline
Magnetic Orientation 
& $J<J_{QPT_1}$& $J>J_{QPT_1}$\\
\hline
$pM_{zy}$ 
& \begin{tabular}[c]{@{}c@{}}
$s$-wave

\end{tabular}& mixed-spin $p_x$-wave\\
\hline
$pM_{zx}$ 
& \begin{tabular}[c]{@{}c@{}}
$s$-wave
\end{tabular}& $s$-wave\\
\hline
\end{tabular}
\end{table}

\begin{table}[h]
\centering
\caption{\label{table:results_summary_competition_py}
Stable pairing symmetries considering the coexistence and competition regimes between singlet $s$-wave and triplet $p_y$-wave superconductivity  for $V=1$ in $J$, obtained from 
Fig.~\ref{fig25}}.
\begin{tabular}{|c|c|c|}
\hline
Magnetic Orientation 
& $J<J_{QPT_2}$& $J>J_{QPT_2}$\\
\hline
$pM_{zy}, pM_{zx}$& \begin{tabular}[c]{@{}c@{}}
$s$-wave, \\
mixed-spin $p_y$-wave

\end{tabular}& equal-spin $p_y$-wave \\\hline
\end{tabular}
\end{table}

\begin{table}[h]
\centering
\caption{\label{table:results_summary_competition2}
Dominant pairing symmetries considering coexistence and competition regimes  between singlet $s$-wave and triplet $p_x$-wave and  $p_y$-wave superconductivity for $V=1$ in $J$, obtained from 
Fig.~\ref{fig26a}.}

\begin{tabular}{|c|c|l|c|}
\hline
Magnetic Orientation 
& $J<J_{QPT_1}$&$J_{QPT_1}<J<J_{QPT_2}$& $J>J_{QPT_2}$\\
\hline
$pM_{zy}$ 
& \begin{tabular}[c]{@{}c@{}}
$s$-wave

\end{tabular}&mixed-spin $p_x$-wave & equal-spin $p_y$-wave \\
\hline
$pM_{zx}$ 
& \begin{tabular}[c]{@{}c@{}}
$s$-wave
\end{tabular}&$s$-wave& equal-spin $p_y$-wave \\
\hline
\end{tabular}
\end{table}

We found that while $s$-wave and mixed-spin $p$-wave symmetries ($S_z=0$) can cooperatively coexist, leading to a mutual enhancement of their critical magnetic couplings $J_c$, the equal-spin triplet states ($S_z=\pm1$) enter into a regime of strong competition with these former pairings. We find two quantum phase transitions: one in the $pM_{zy}$ magnetic orientation where mixed-spin $p_x$-wave surpasses $s$-wave superconductivity, and another in both the $pM_{zy}$ and the $pM_{zx}$ magnetic orientations, separating the competing phases that mutually exclude each other. Both of these transitions are induced by the magnetic coupling of the $p$-wave magnet, highlighting them as platforms for superconducting symmetry selection. Additionally, our study of proximity effects indicates that a stable singlet $s$-wave background can act as a symmetry anchor, effectively ``softening'' the pair-breaking effects of the magnet and allowing $p_x$-wave triplets to survive at magnetic intensities above their individual limits. These findings suggest that heterostructures involving $p$-wave magnets like $\text{CeNiAsO}$ or $\text{Mn}_3\text{GaN}$ offer a tunable platform for engineering robust triplet superconductivity. Future experimental investigations using tunneling spectroscopy or Josephson junction transport could verify these predicted phase transitions and the stabilization of the mixed-spin triplet sector.

\clearpage

\section*{Acknowledgement}

This work was supported by FCT under research unit UID/00618 - Center for Theoretical and Computational Physics, and research unit UID/04540 - Center of Physics and Engineering of Advanced Materials, and the contract LA/P/0095/2020, LaPMET, Laboratory of Physics for Materials and Emerging Technologies.


\end{document}